# FIRST POLARIMETRIC MEASUREMENTS AND MODELING OF THE PASCHEN-BACK EFFECT IN CaH TRANSITIONS

S. V. Berdyugina[1] and D. M. Fluri
Institute of Astronomy, ETH Zurich, 8092, Zurich, Switzerland; sveta@astro.phys.ethz.ch, fluri@astro.phys.ethz.ch

R. Ramelli and M. Bianda[2]
Istituto Ricerche Solari Locarno, Via Patocchi, CH-6605 Locarno-Monti, Switzerland; ramelli@irsol.ch, mbianda@irsol.ch

and

D. Gisler and J. O. Stenflo[3]
Institute of Astronomy, ETH Zurich, 8092, Zurich, Switzerland; gisler@astro.phys.ethz.ch, stenflo@astro.phys.ethz.ch

## ABSTRACT

We report the first spectropolarimetric observations and modeling of CaH transitions in sunspots. We have detected strong polarization signals in many CaH lines from the $A$–$X$ system, and we provide the first successful fit to the observed Stokes profiles using the previously developed theory of the Paschen-Back effect in arbitrary electronic states of diatomic molecules and polarized radiative transfer in molecular lines in stellar atmospheres. We analyze the CaH Stokes profiles together with quasi-simultaneous observations in TiO bands and conclude that CaH provides a valuable diagnostic of magnetic fields in sunspots, starspots, cool stars, and brown dwarfs.

*Subject headings:* magnetic fields — molecular processes — polarization — sunspots

*Online material:* color figures

## 1. INTRODUCTION

CaH is one of the most important astrophysical molecules. Two red-band systems of CaH, $A\ ^2\Pi$–$X\ ^2\Sigma^+$ and $B\ ^2\Sigma^+$–$X\ ^2\Sigma^+$, were first detected in spectra of M dwarfs by Fowler (1907) and almost immediately after that identified in the sunspot spectrum by Olmsted (1908). Later it was established that the formation of CaH in stellar atmospheres is strongly influenced by gas pressure; thus, CaH bands, as well as other metal hydride bands, can be used as a luminosity indicator of cool stars (Öhman 1934; Mould 1976; Mould & Wallis 1977 ; Barbuy et al. 1993). Absorption in CaH bands is also an important opacity source in brown dwarfs and can be used for studying these substellar objects (e.g., Burrows et al. 2001). The astrophysical value of CaH has motivated a series of laboratory spectroscopic studies of the molecule (e.g., Berg & Klynning 1974a, 1974b; Martin 1984; Weck et al. 2003; Shayesteh et al. 2004; Steimle et al. 2004). Furthermore, because of its remarkable magnetic properties, CaH was the first molecule having been cooled down to nanokelvin temperatures to produce molecular Bose-Einstein condensation in a magnetic trap (Weinstein et al. 1998).

A laboratory study of the red CaH bands in a magnetic field started immediately after Hill (1929) had formulated the first theory of the Zeeman and Paschen-Back effects in molecular doublet states for the case intermediate between Hund's cases (*a*) and (*b*). This theory explained well Zeeman shifts of transitions in the $A$–$X$ system with smaller rotational numbers. Transitions between higher rotational levels, as well as those in the $B$–$X$ system, were found to be affected by the rotational perturbation between the excited states $A\ ^2\Pi$ and $B\ ^2\Sigma^+$ (Watson & Bender 1930; Watson 1932; Cunningham & Watson 1933). Despite the progress in explaining Zeeman shifts of CaH transitions, no spectropolarimetric measurements were made to date neither in laboratories, nor in astrophysics. The only report of possible Zeeman broadening of CaH lines in sunspots was given by Mallia (1970).

A study of solar and stellar magnetic fields with molecular spectropolarimetry is a new and rapidly developing field pioneered by Berdyugina et al. (2000). Molecules are preferably formed in cooler environments and thus bear the information on their physical conditions, including the magnetic field. An overview of magnetic properties of astrophysically important diatomic molecules (Berdyugina & Solanki 2002; Berdyugina et al. 2003, 2005; Asensio Ramos & Trujillo Bueno 2006) has indicated a great potential of molecular spectropolarimetry for studying cosmic magnetic fields. CaH lines are especially useful diagnostic of sunspots with the umbral temperature of 4000–4250 K, at which absorption in CaH is significantly stronger than that in TiO (Berdyugina et al. 2003). Moreover, because of this property, CaH lines may be more readily detected in starspots on active G–K stars than TiO lines. A simultaneous analysis of CaH and TiO lines can therefore provide valuable constraints for temperature, pressure, and magnetic field in the atmosphere of sunspots, starspots, and cool stars.

Here we report the first spectropolarimetric measurements and modeling of CaH Stokes profiles. We detect strong polarization features in many CaH lines from the $A$–$X$ system observed in sunspots and successfully fit their Stokes profiles using the previously developed theory of the Paschen-Back effect (PBE) in arbitrary electronic states of diatomic molecules (Berdyugina et al. 2005) and polarized radiative transfer of molecular lines in the stellar atmospheres (Berdyugina et al. 2003). We therefore confirm that CaH is a valuable diagnostic of magnetic fields in cool astrophysical sources. We describe our polarimetric observations in § 2. Then, in § 3, we discuss the PBE in the CaH $A$–$X$ system and present calculations of the Zeeman patterns. Modeling Stokes profiles of the CaH lines and a comparison with observations are presented in § 4. Finally, we summarize our results and draw the conclusions in § 5.

[1] Astronomy Division, University of Oulu, P.O. Box 3000, FIN-90014 Oulu, Finland.
[2] Institute of Astronomy, ETH Zurich, 8092, Zurich, Switzerland.
[3] Faculty of Mathematics and Science, University of Zurich, Zurich, Switzerland.

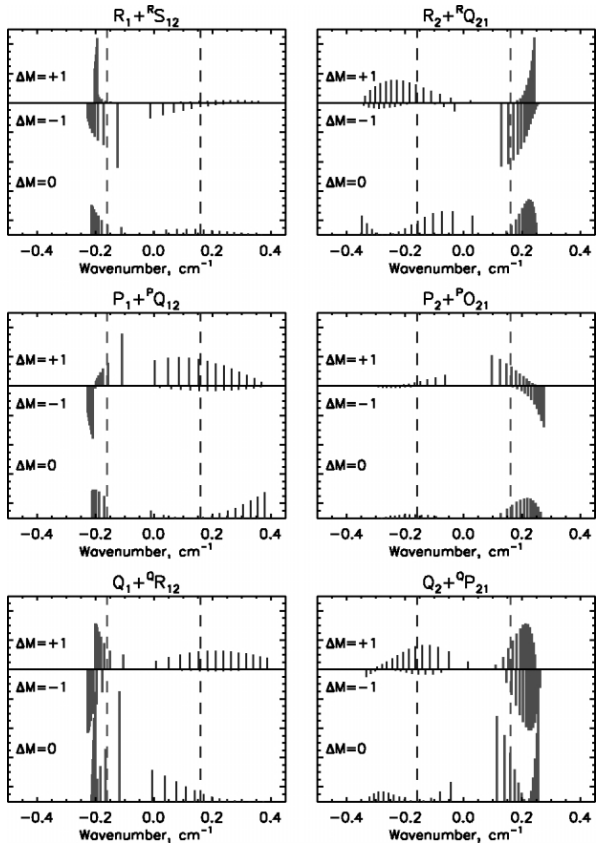

Fig. 1.—CaH magnetic patterns for different rotational branches arising from the lower rotational levels with $N = 7$ at the magnetic field strength of 3 kG. The Zeeman components shown by thick and thin sticks correspond to the main and satellite lines, respectively. The components with $\Delta M = -1$ are plotted downward for clarity. Vertical dashed lines indicate the positions of the main and satellite lines in zero magnetic field with respect to the doublet center, which is at 0.0 cm$^{-1}$. [*See the electronic edition of the Journal for a color version of this figure.*]

## 2. OBSERVATIONS

The spectropolarimetric observations were carried out at the Gregory Coudé Telescope of the Istituto Ricerche Solari Locarno (IRSOL) with the Zurich IMaging POLarimeter (ZIMPOL; Povel 2001). Due to high modulation frequency of 42 kHz (which is much higher than typical seeing frequencies), ZIMPOL allows us to obtain measurements that are unaffected by seeing induced spurious effects. Calibration included measurements of the polarimetric efficiency, dark current, flat field, and instrumental polarization (Ramelli et al. 2005).

We observed a large spot in the active region NOAA AR 822, which passed across the solar disk on 2005 November 17–24. The slit was placed across the umbra's darkest part, and the spectra were recorded from the umbra, both sides of the penumbra, and the photosphere. The ZIMPOL system allowed simultaneous measurements of Stokes $IQV$ or $IUV$. Both sequences were recorded for each wavelength region. Calibrations were performed as explained by Gandorfer et al. (2004). Measurements of the instrumental polarization are used to correct the data as shown by Ramelli et al. (2005). The stray light introduced by the instrumentation is estimated to be about 5%. To account for this effect, calculated spectra were diluted with 5% of the synthetic photospheric spectrum.

To increase the signal-to-noise ratio, we averaged spectra along the slit area covering the central part of the umbra. A correction for the telluric line absorption was applied using the transmission spectrum from the NSO atlas by Wallace et al. (1998).

Several spectral regions with strong CaH lines were observed on different dates. Here we analyze two regions centered at 693.9 and 697.2 nm, which were observed on 2005 November 17–21. The regions include lines mainly from the $A$–$X$ (0, 0) band, which were expected to show strong polarization signals. Since CaH lines are blended with numerous weak TiO lines from the $\gamma$-system, two other regions centered at 705.7 and 759.0 nm were also observed. They include the TiO (0, 0) and (0, 1) band heads of the $\gamma$-system. The first band is known to show strong polarization signatures (Berdyugina et al. 2000), and the second one belonging to the same electronic system is expected to have similar magnetic sensitivity. These are the first polarimetric observations of CaH lines and of the TiO $\gamma(0, 1)$ band.

## 3. PASCHEN-BACK EFFECT IN THE CaH $A$–$X$ SYSTEM

Recently, we presented a general numerical approach to the molecular PBE that is valid for terms of any multiplicity and accounts for interactions of all rotational levels in an electronic state intermediate between Hund's cases (*a*) and (*b*) (Berdyugina et al. 2005). This allows us to develop novel diagnostic tools for studying magnetic fields in astrophysical objects. CaH transitions are obviously among those of the first choice.

The CaH $A\,^2\Pi$–$X\,^2\Sigma^+$ band system is observed in the wavelength region 660–760 nm. The ground state $X\,^2\Sigma^+$ is well described by a pure Hund's case (*b*), since the projection of the electronic orbital momentum on the internuclear axis $\Lambda$ is zero. Small spin-rotation interaction results in the splitting of each rotational level onto two levels of the fine structure. The spin-rotation coupling constant of 0.0436 cm$^{-1}$ (Martin 1984) indicates that significant perturbations are already seen at magnetic fields stronger than 100 G, and the complete PBE in the lowest rotational level occurs at 470 G. The excited $A\,^2\Pi$ state is intermediate between Hund's cases (*a*) and (*b*). The relatively large spin-orbit coupling constant of $\sim 79$ cm$^{-1}$ and rotational constant of $\sim 4.3$ cm$^{-1}$ (Martin 1984) suggest that lower rotational levels are closer to Hund's case (*a*), while higher levels approach Hund's case (*b*) due to gradual spin uncoupling from the internuclear axis with increasing rotation. The PBE in this electronic state becomes noticeable only at magnetic fields stronger than 20 kG and in higher rotational levels.

Rotational levels in the $A\,^2\Pi$ vibrational states with $v \geq 1$ perturb the nearby levels of the $B\,^2\Sigma^+$ state and lend some of their properties to the $B$ state. This perturbation also results in $\Lambda$-doubling of the $A$ state. Accounting for this perturbation should result in different magnetic properties of transitions of different parities. This is left out in the present consideration and will be investigated in the future.

Using molecular constants determined by Berg & Klynning (1974a) and Martin (1984) and the PBE theory as described by Berdyugina et al. (2005), we calculated wavelength shifts and strengths of Zeeman components for all CaH lines in the observed spectral regions. Since the PBE results in the appearance of forbidden transitions (O and S branches), they were also included in the calculation. Their zero-field wavelengths were predicted using those of allowed transitions measured in the laboratory by S. Davis (2000, private communication). Examples of Zeeman patterns for different main and satellite rotational branches are shown in Figure 1. Note that the PBE noticeably alters relative strengths of transitions with different $\Delta M$, resulting in strongly asymmetric patterns and wavelength shifts of the line center depending on the magnetic field strength. Also, the total strength of the main branch lines decreases, while that of satellite and forbidden lines increases.

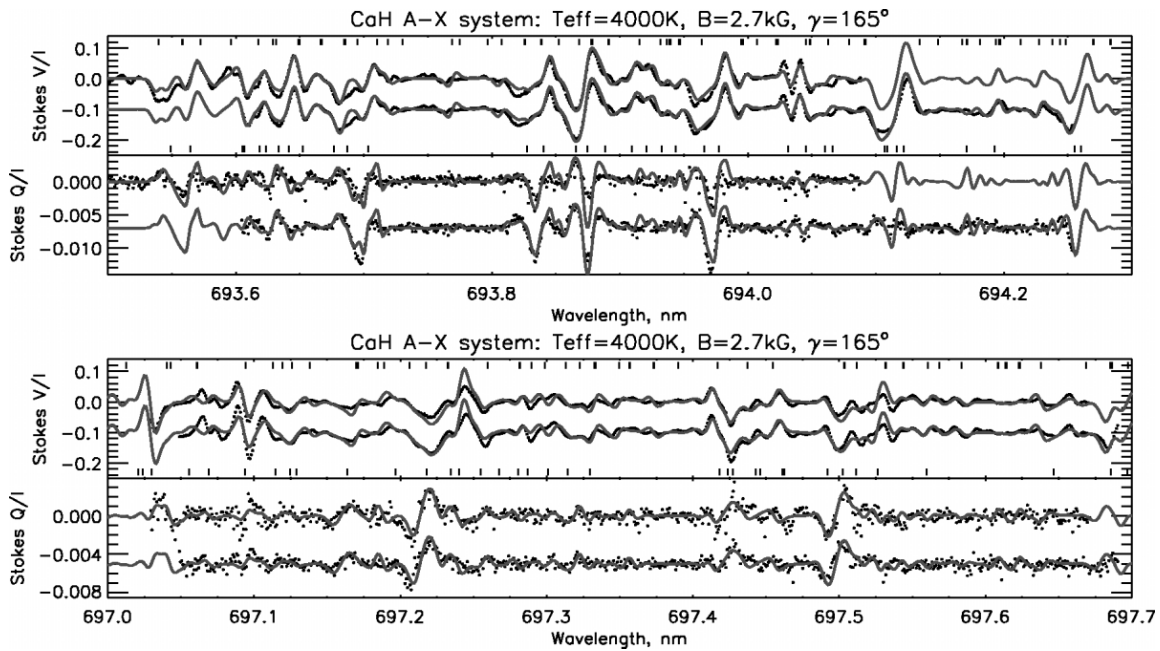

FIG. 2.—CaH Stokes profiles in the sunspot umbra. Dots represent observations: two different dates for each region, one being shifted in vertical direction for clarity. Solid lines are synthetic spectra. The 693.9 nm region was observed on 2005 November 18 and 19, the other one on 2005 November 17 and 19. In the top panels, vertical dashes indicate positions of TiO lines (*top edge*) and CaH lines (*bottom edge*). The effective temperature of the equilibrium model and the magnetic field strength and angle to the line of sight indicated above the plots have been chosen to fit reasonably both pairs of observations. [*See the electronic edition of the Journal for a color version of this figure.*]

### 4. STOKES PROFILES

The polarized radiative transfer in all four Stokes parameters for molecular lines in stellar atmospheres was solved according to the description given by Berdyugina et al. (2003). The Stokes parameter synthesis was done using equilibrium stellar atmosphere models by Kurucz (1993).

In the two CaH spectral regions (Fig. 2) the strongest lines are from the least perturbed vibrational band (0, 0) of the CaH $A$–$X$ system. A number of weaker lines from the (1, 1) and (2, 2) bands are also present. In addition to the CaH lines, many TiO lines from the $\gamma$-system were included in the spectrum synthesis. These lines also show a remarkable Zeeman effect (Berdyugina et al. 2000), but here they originate from higher vibrational states and appear weaker than CaH lines. Also, a few atomic and MgH $B'$–$X$ blends were included in the line list. Despite the complexity of various line contributions, there is a remarkable resemblance between the observed and calculated

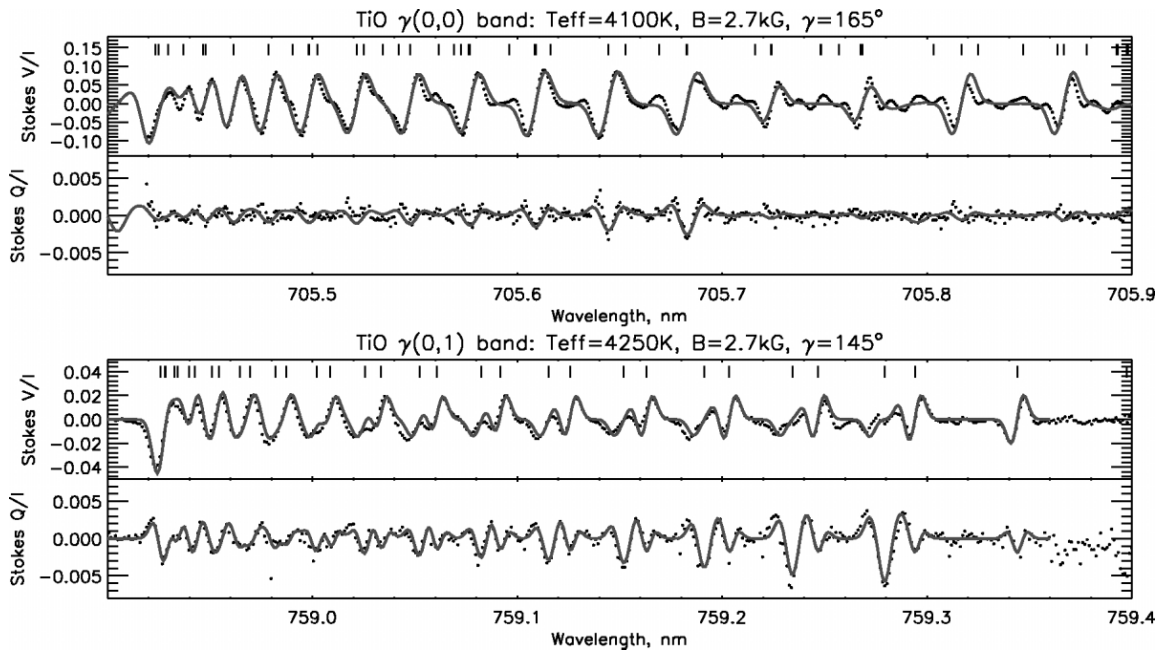

FIG. 3.—Same as Fig. 2, but for TiO Stokes profiles. The 705.7 nm region was observed on 2005 November 19, and the 759.0 nm on 2005 November 21. [*See the electronic edition of the Journal for a color version of this figure.*]

Stokes profiles (Fig. 2). This implies that our PBE calculation for the CaH $A$–$X$ system is adequate for transitions with minimal perturbation by the $B$ state. Small discrepancies in the fits are mainly due to missing blends but can also be due to the neglected perturbation. The peculiar Stokes profiles caused by the PBE are extremely sensitive to the magnetic field strength.

The regions with the TiO bands are only slightly blended by other lines. At sunspot magnetic fields the TiO transitions are well described by the Zeeman regime calculations for the intermediate Hund's case ($a$–$b$) (Berdyugina & Solanki 2002; Berdyugina et al. 2003). Prominent polarization signatures are observed in both bands (Fig. 3): Stokes $V$ is the strongest for the near disk center observation (705.7 nm region), while Stokes $Q$ becomes comparable to Stokes $V$ for the off center observation (759.0 nm region).

The high sensitivity to the temperature and magnetic field strength of the CaH and TiO Stokes profiles allowed us to constrain these parameters in the umbra. Atmosphere models with somewhat different effective temperatures were employed to obtain good fits to the CaH and TiO profiles in different spectral regions. This indicates that simple equilibrium models are not able to reproduce the complex structure of sunspot umbrae revealed by molecular lines, and a more sophisticated, multicomponent analysis based on inversion techniques is needed. Our analysis suggests that simultaneous spectropolarimetry in CaH and TiO lines with high spatial resolution would be essential for revealing the internal structure of the sunspot umbra and for understanding its heating mechanisms.

## 5. CONCLUSIONS

First spectropolarimetric observations of the CaH $A\,^2\Pi$–$X\,^2\Sigma^+$ system have revealed strong polarization signatures in many lines. Peculiar Stokes profiles observed in sunspots are caused mainly by the Paschen-Back effect on the fine structure of the ground state which significantly perturbs the internal molecular structure at fields stronger than $\sim$100 G.

Our advanced calculation of the molecular PBE combined with the full Stokes radiative transfer provided the first successful fit to the observed CaH Stokes profiles. This opens a new opportunity for studying magnetic fields in sunspots and starspots and also in very cool stars and brown dwarfs.

We are thankful to Sumner Davis for providing laboratory wavelengths of CaH transitions. The observations at IRSOL are possible thanks to the financial support that has been provided by the canton of Ticino, the city of Locarno, ETH Zurich, and the Fondazione Carlo e Albina Cavargna. S. V. B. acknowledges the EURYI award from the ESF, SNF grants PE002-104552 and 200021-103696, and ETH research grant TH-2/04-3.